\documentclass[journal,twoside,web]{ieeecolor}
\usepackage{tmi}
\usepackage{cite}
\usepackage{amsmath,amssymb,amsfonts}
\usepackage{graphicx}
\usepackage{textcomp}
\usepackage{algorithm}
\usepackage{algpseudocode}

\usepackage{makecell}
\usepackage{colortbl}
\usepackage{xcolor}
\usepackage{booktabs}
\usepackage{multirow}

\def\BibTeX{{\rm B\kern-.05em{\sc i\kern-.025em b}\kern-.08em
    T\kern-.1667em\lower.7ex\hbox{E}\kern-.125emX}}
\begin{document}
\title{Diffusion Probabilistic Priors for Zero-Shot Low-Dose CT Image Denoising}
\author{Xuan Liu, Yaoqin Xie, Jun Cheng, Songhui Diao, Shan Tan, \IEEEmembership{Member, IEEE}, and Xiaokun Liang, \IEEEmembership{Member, IEEE}
\thanks{This work is partly supported by grants from the National Natural Science Foundation of China (62071197, 82202954, U20A20373, U21A20480, 12126608) and the Chinese Academy of Sciences Special Research Assistant Grant Program.(\textit{Corresponding author: Shan Tan and Xiaokun Liang.})}
\thanks{Xuan Liu is with the
School of Artificial Intelligence and Automation, Huazhong University of Science and Technology, Wuhan 430074, China and the Institute of Biomedical and Health Engineering, Shenzhen Institutes of Advanced Technology, Chinese Academy of Sciences, Shenzhen 518055, China (e-mail: liuxuan99@hust.edu.cn).}
\thanks{Jun Cheng and Shan Tan are with the
School of Artificial Intelligence and Automation, Huazhong University of Science and Technology, Wuhan 430074, China (e-mail: d202180985@hust.edu.cn, shantan@hust.edu.cn).}
\thanks{Yaoqin Xie, Songhui Diao, and Xiaokun Liang are with the Institute of Biomedical and Health Engineering, Shenzhen Institutes of Advanced Technology, Chinese Academy of Sciences, Shenzhen 518055, China (e-mail: yq.xie@siat.ac.cn, sh.diao@siat.ac.cn, xk.liang@siat.ac.cn).}
}

\maketitle

\begin{abstract}
Denoising low-dose computed tomography (CT) images is a critical task in medical image computing. Supervised deep learning-based approaches have made significant advancements in this area in recent years. However, these methods typically require pairs of low-dose and normal-dose CT images for training, which are challenging to obtain in clinical settings. Existing unsupervised deep learning-based methods often require training with a large number of low-dose CT images or rely on specially designed data acquisition processes to obtain training data. To address these limitations, we propose a novel unsupervised method that only utilizes normal-dose CT images during training, enabling zero-shot denoising of low-dose CT images. Our method leverages the diffusion model, a powerful generative model. We begin by training a cascaded unconditional diffusion model capable of generating high-quality normal-dose CT images from low-resolution to high-resolution. The cascaded architecture makes the training of high-resolution diffusion models more feasible. Subsequently, we introduce low-dose CT images into the reverse process of the diffusion model as likelihood, combined with the priors provided by the diffusion model and iteratively solve multiple maximum a posteriori (MAP) problems to achieve denoising. Additionally, we propose methods to adaptively adjust the coefficients that balance the likelihood and prior in MAP estimations, allowing for adaptation to different noise levels in low-dose CT images. We test our method on low-dose CT datasets of different regions with varying dose levels. The results demonstrate that our method outperforms the state-of-the-art unsupervised method and surpasses several supervised deep learning-based methods. Codes are available in \underline
{https://github.com/DeepXuan/Dn-Dp}.
\end{abstract}


\section{Introduction}
\label{sec:introduction}
\IEEEPARstart{L}{ow-dose} computed tomography (CT) has garnered increasing attention due to its significant reduction in radiation dose. While low-dose CT reduces the risk of patient exposure to X-rays, it often results in poor-quality reconstructed images with noticeable noise, which can hinder accurate diagnosis. Various algorithms have been developed to enhance the image quality of low-dose CT, including sinogram filtering~\cite{balda2012ray}, iterative reconstruction~\cite{sun2015iterative}, and image denoising~\cite{chen2017low}. Among these methods, low-dose CT image denoising is the most widely used, as it does not involve the reconstruction process and can be applied to different CT imaging systems.

In the past decade, deep neural networks, fueled by the rapid advancement of deep learning techniques, have emerged as a powerful tool for low-dose CT image denoising, achieving remarkable success. Typically, deep learning-based methods learn the end-to-end mapping from low-dose to normal-dose CT images in a supervised manner. Initially, convolutional neural networks (CNNs) were the most commonly employed for low-dose CT image denoising~\cite{chen2017low}. To improve the visual quality of denoised images, generative adversarial networks (GANs) have also been utilized in low-dose CT image denoising~\cite{yang2018low,yi2018sharpness}. Additionally, transformer-based networks further enhance the performance of low-dose CT image denoising~\cite{wang2023ctformer,li2022transformer}. 

However, the success of supervised deep learning algorithms heavily relies on a large amount of paired training data~\cite{liu2022learning}. Unfortunately, acquiring such data in clinical settings is challenging due to the adherence to the "as low as reasonably achievable" principle~\cite{brenner2007computed}. Therefore, it is crucial to develop methods that can exploit the high performance of deep neural networks without requiring a significant amount of labeled data.

Some existing approaches have attempted to address this problem using different strategies. One category of methods rely on training data obtained through special acquisition schemes. Yuan et al. proposed Half2Half, which trained a neural denoiser using paired half-dose CT images~\cite{yuan2020half}. Wu et al. developed a method to train low-dose CT image denoising networks using pairs of images reconstructed from odd and even projections~\cite{2019Consensus}. Another category of methods further relax the conditions for training data but still requires a large number of low-dose CT images for training. For example, Du et al. proposed to learn invariant representation from noisy images and reconstruct clean observations~\cite{du2020learning}. Liu et al. employed an invertible neural network to simulate paired normal-dose and low-dose CT images with unpaired training~\cite{liu2022learning}. Niu et al. introduced a similarity-based unsupervised deep denoising method to handle correlated noise in CT images~\cite{niu2022noise}.

Unlike the aforementioned unsupervised methods, the method we propose in this paper does not require paired normal-dose and low-dose CT images, nor does it need any low-dose CT images as a training set. We utilize only normal-dose CT images to train a denoising diffusion probabilistic model\cite{ho2020denoising,song2020denoising} (referred to as the diffusion model hereafter) then leverage the prior information embedded in the pre-trained diffusion model to achieve zero-shot low-dose CT image denoising. Diffusion models define a constant forward process and achieve image generation by iteratively solving the reverse process. Not only do diffusion models achieve state-of-the-art generation results, but they also provide an analytical representation of the data distribution during the generation process, which is not achievable by models such as GANs. Thanks to the powerful generative and representational capabilities of the diffusion model, some existing works have explored the application of its priors in downstream tasks such as image reconstruction and restoration~\cite{kawar2021snips,kawar2022denoising,song2022solving,peng2022towards}. However, most of these methods either neglect the presence of noise~\cite{peng2022towards} or only consider simple noise scenarios~\cite{kawar2021snips,kawar2022denoising,song2022solving}. In our approach, we first propose incorporating the diffusion priors into the maximum a posteriori (MAP) framework to address low-dose CT image denoising. Furthermore, we introduce a strategy to dynamically adjust the weights of the prior and likelihood to handle complex low-dose CT noise, which exhibits different levels across different CT slices.


To begin, we train a diffusion model using normal-dose CT images. Previous studies have highlighted the challenges in directly generating high-resolution images of size 512$\times$512. A cascaded generation approach is employed to address this issue~\cite{karras2017progressive,karras2019style,ho2022cascaded}. Specifically, we generate low-resolution images from random noise and then produce high-resolution images of size 512$\times$512 by conditioning on the low-resolution images. Then, we propose an algorithm that incorporates the pre-trained diffusion model into the MAP denoising framework. In each iteration of the diffusion model's generation process, we introduce low-dose CT images and solve a MAP estimation problem, ensuring a good likelihood between the generated images and the input low-dose CT images. Ultimately, the diffusion model can produce high-quality counterparts of the input low-dose CT images, achieving low-dose CT image denoising. Moreover, considering that different low-dose CT slices exhibit varying noise levels, we design two adaptive strategies to refine the coefficients in the MAP estimations, which balance the likelihood and prior. By resuming the denoising process at an intermediate timestep with the refined coefficients, our algorithm can adaptively deal with low-dose CT images with varying noise levels.

Our contributions can be summarized as follows: (1) We propose to train diffusion models on large-scale CT image datasets in a cascaded manner, which can progressively generate realistic low-resolution and high-resolution normal-dose CT images from random noise. (2) We propose an algorithm for fully unsupervised zero-shot low-dose CT image denoising with diffusion priors, by solving multiple MAP problems in the reverse process of the diffusion model. (3) We introduce two adaptive strategies for the coefficients in the MAP estimations, balancing the likelihood and prior to handle different levels of low-dose CT noise. (4) Our method demonstrates excellent results in terms of both visual effects and metrics, outperforming the state-of-the-art unsupervised method and even surpassing several supervised deep learning-based methods.

\section{Method}
\subsection{NDCT Image Generation with Diffusion Models}
We employ a two-step cascade of diffusion models to generate normal-dose CT images. The first model is an unconditional diffusion model that generates low-resolution normal-dose CT images from random noise. The second model is a conditional diffusion model that takes the low-resolution images as conditions to generate high-resolution normal-dose CT images of size 512$\times$512. The overall model is unsupervised, as it only uses normal-dose CT images without labels for training.
\begin{figure*}[t]
	\centering\includegraphics[width=0.95\textwidth]{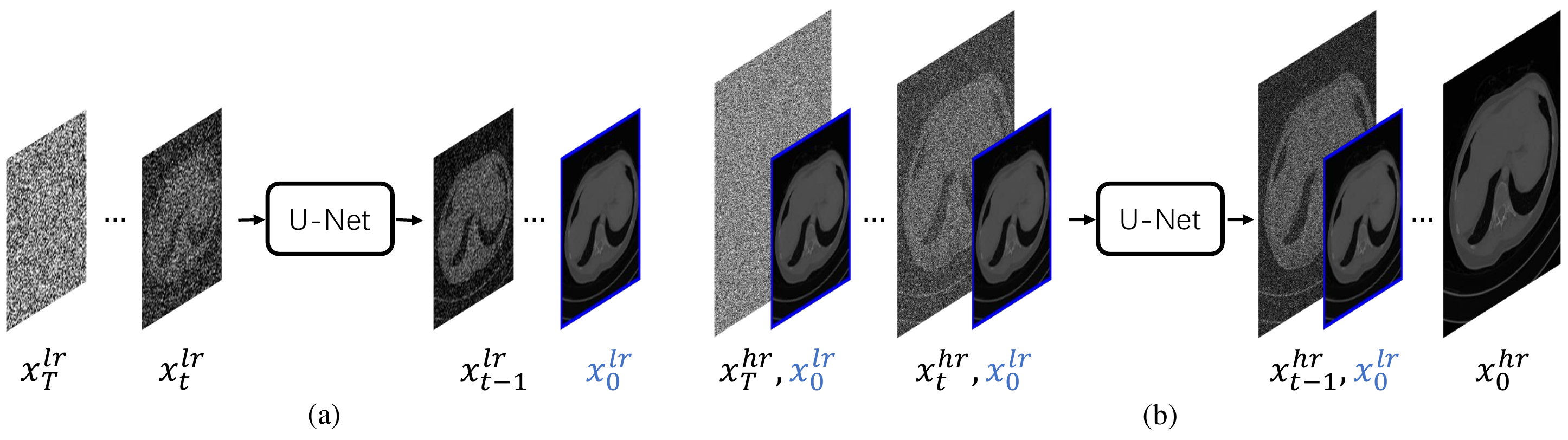}
	\caption{Overall generation process of pretrained diffusion models. (a) Illustrates the generation of a low-resolution CT image $x_0^{lr}$ from random noise $x_T^{lr}$. (b) Depicts the generation of the high-resolution CT image using $x_0^{lr}$ as a condition.} \label{fig1}
\end{figure*}
\subsubsection{Unconditional Diffusion Model}
Given a set of data $x_0\sim q\left(x_0\right)$, diffusion models define a fixed forward process and aim to reverse it to realize image generation from random noise. The forward process $q\left(x_{1: T} \mid x_0\right)$ is defined as a Markov chain, where the probability distribution at each timestep depends only on the previous timestep. Specifically, we have:
\begin{equation}
	\begin{split}
		q\left(x_{1: T} \mid x_0\right)&:=\prod_{t=1}^T q\left(x_t \mid x_{t-1}\right), q\left(x_t \mid x_{t-1}\right)\\
		&:=\mathcal{N}\left(x_t ; \sqrt{1-\beta_t} x_{t-1}, \beta_t I\right),\label{eq1}
	\end{split}
\end{equation}
where $x_t$ denotes the data at timestep $t$, $\mathcal{N}\left(\cdot; \mu, \Sigma \right)$ denotes a Gaussian distribution with mean $\mu$ and covariance matrix $\Sigma$, and ${\beta_1, \ldots, \beta_T}$ is a fixed variance schedule. The forward process progressively adds noise to $x_0$~\cite{song2020denoising}. Based on (\ref{eq1}), we can derive the following two conditional distributions:

\begin{equation}
	q\left(x_t \mid x_0\right)=\mathcal{N}\left(x_t ; \sqrt{\bar{\alpha}_t} x_0,\left(1-\bar{\alpha}_t\right) I\right),\label{eq2}
\end{equation}

\begin{equation}
	q\left(x_{t-1} \mid x_t, x_0\right)=\mathcal{N}\left(x_{t-1} ; \tilde{\mu}_t\left(x_t, x_0\right), \tilde{\beta}_t I\right),\label{eq3}
\end{equation}

where,
	\begin{align}
		&\bar{\alpha}_t:=\prod_{i=1}^t (1-\beta_i),\\
		&\tilde{\mu}_t\left(x_t, x_0\right):=\frac{\sqrt{\bar{\alpha}_{t-1}} \beta_t}{1-\bar{\alpha}_t} x_0+\frac{\sqrt{\alpha_t}\left(1-\bar{\alpha}_{t-1}\right)}{1-\bar{\alpha}_t} x_t, \\
		&\tilde{\beta}_t:=\frac{(1-\bar{\alpha}_{t-1})}{1-\bar{\alpha}_t} \beta_t. 
	\end{align}

The reverse process is defined as a Markov chain with learnable parameters $\theta$, 
\begin{equation}
	p_\theta\left(x_{0: T}\right):=p\left(x_T\right) \prod_{t=1}^T p_\theta\left(x_{t-1} \mid x_t\right),
\end{equation}
where,
	\begin{align}
		&p\left(x_T\right)=\mathcal{N}\left(x_T ; 0, I\right),\\
		&p_\theta\left(x_{t-1} \mid x_t\right):=\mathcal{N}\left(x_{t-1} ; \mu_\theta\left(x_t, t\right), \sigma^2_{t}I\right).\label{eq4} 
	\end{align}  
Here, $\mu_\theta\left(x_t, t\right)$ is generally predicted by U-nets, and $\sigma^2_{t}$ is defined as $\frac{1-\bar{\alpha}_{t-1}}{1-\bar{\alpha}_{t}}\beta_t$. The training loss is derived by minimizing the variational bound of the negative log likelihood $-\log p_\theta\left(x_0\right)$~\cite{ho2020denoising}, 

\subsubsection{Conditional Diffusion Model}
Given a dataset of image pairs $(x_0, c)$, a conditional diffusion model maintains the same forward process as an unconditional model, where the forward process is independent of $c$ and follows (\ref{eq1}). However, the reverse process takes $c$ as an additional condition, which can be defined as $p_\theta\left(x_{0: T} \mid c\right):=p\left(x_T\right) \prod_{t=1}^T p_\theta\left(x_{t-1} \mid x_t, c\right)$, where 
\begin{equation}
	p_\theta\left(x_{t-1} \mid x_t, c\right):=\mathcal{N}\left(x_{t-1} ; \mu_\theta\left(x_t, t, c\right), \sigma^2_{t}\right).\label{eq6}
\end{equation}    

\subsubsection{Cascaded Diffusion Model for NDCT Image Generation}  
We propose to use a cascaded framework for generating high-resolution normal-dose CT images, as previous studies have shown the challenges of directly generating high-resolution images~\cite{karras2017progressive,karras2019style,ho2022cascaded}.
Given a set of normal-dose CT images $x_0^{hr} \in \mathbb{R}^{512\times512}$, we obtain image pairs $(x_0^{hr}, x_0^{lr})$ by downsampling $x_0^{hr}$ by a factor of $k$, where $x_0^{lr} \in \mathbb{R}^{\frac{512}{k}\times\frac{512}{k}}$. We first define a diffusion model with forward process parameters $\{\bar{\alpha}^{lr}_t\}_{t=1,\ldots,T}$ and reverse process parameters $\theta^{lr}$, which is used to generate low-resolution normal-dose CT images from random noise. The second diffusion model is a conditional diffusion model that takes a low-resolution CT image as a condition and realizes normal-dose CT image super-resolution. Its forward process parameters are $\{\bar{\alpha}^{sr}_t\}_{t=1,\ldots,T}$ and reverse process parameters are $\theta^{sr}$. By cascading the two diffusion models, a low-resolution normal-dose CT image is first generated from random noise, and then the generated low-resolution CT image is used as a condition in the super-resolution diffusion model to generate a high-resolution normal-dose CT image. The overall generation process is illustrated in Fig. \ref{fig1}. 

\subsection{Low-dose CT Image Denoising with Diffusion Priors}

In this section, we propose an algorithm, called Dn-Dp (Denoising with Diffusion Prior), for low-dose CT image denoising using diffusion priors. We first describe the algorithm in a general case and then apply it to the specific problem of low-dose CT image denoising by utilizing our pre-trained diffusion models.

\begin{figure*}[t]
	\centering\includegraphics[width=0.95\textwidth]{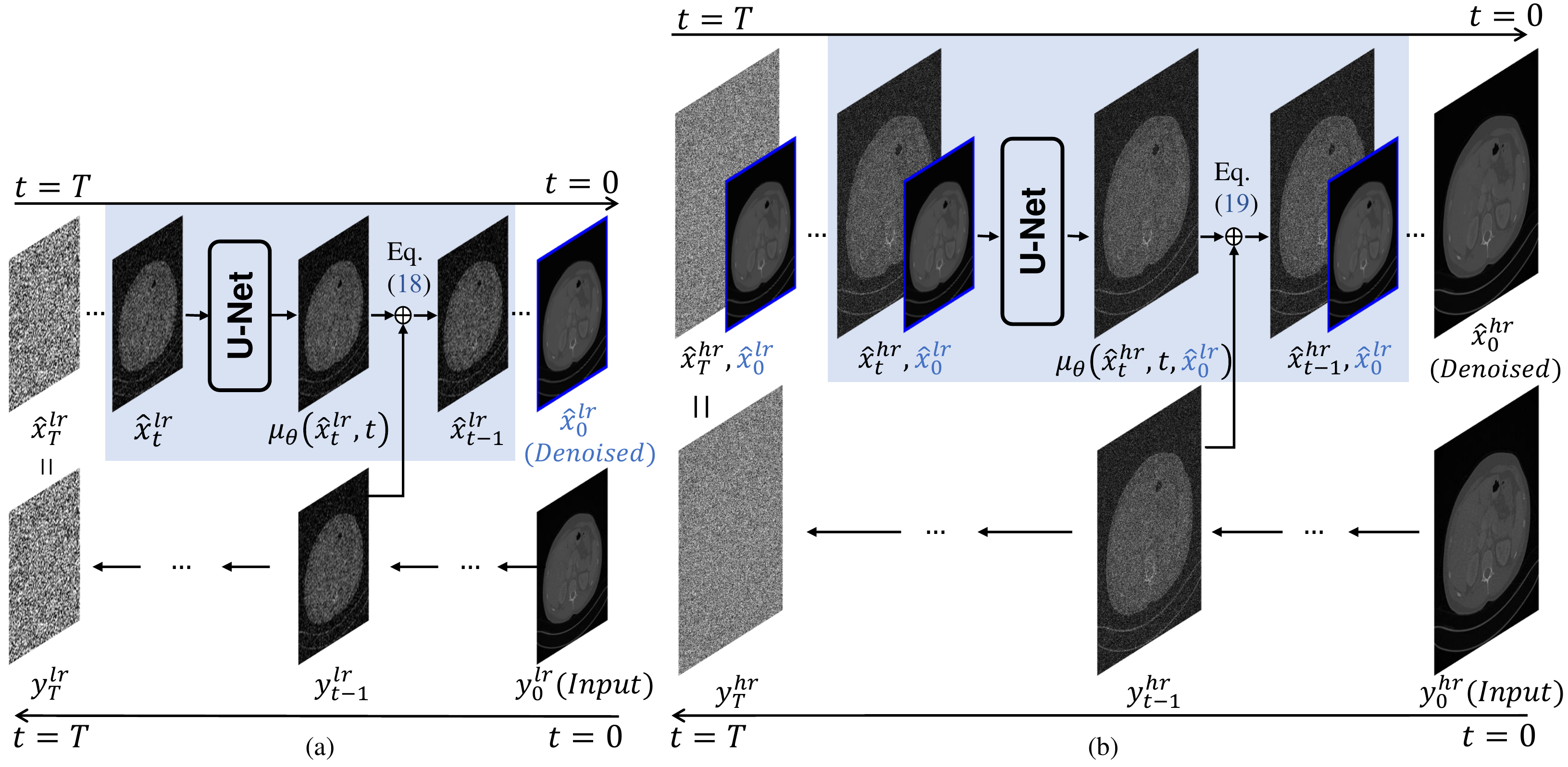}
	\caption{Proposed workflow for low-dose CT image denoising. (a) Presents the denoising process of the low-resolution image $y_0^{lr}$ based on an unconditional diffusion model. (b) Illustrates the denoising process of the high-resolution image $y_0^{hr}$ using the denoising result from (a) as a condition. The blue blocks represent one iteration of our denoising algorithms.} \label{fig2}
\end{figure*} 

\subsubsection{Iterative Denoising with Diffusion Priors}

In image denoising problems, the corruption can be modeled as follows:
\begin{equation}
    y_0 = x_0 + n, \label{eq8}
\end{equation}
where $y_0$ is the observed noisy image and $x_0$ is the underlying clean image we aim to estimate. To achieve this, we leverage a well-trained diffusion model that accurately captures the prior distribution of $x_0$ using forward process parameters $\{\bar{\alpha}_t\}_{t=1,\ldots,T}$ and reverse process parameters $\theta$. The goal is to effectively denoise $y_0$ using the diffusion prior. The concrete method is outlined below.

In diffusion models, the image $x_t$ at timestep $t=1, \ldots, T$ can be generated from $x_0$ based on the forward process:
\begin{equation}
    x_t = \sqrt{\bar{\alpha}_t} x_0 + \sqrt{1-\bar{\alpha}_t}\epsilon_t, \label{eq9}
\end{equation}
where $\epsilon_t\sim\mathcal{N}\left(0, I\right)$ is drawn from a normal distribution. To introduce the likelihood from the input noisy image $y_0$ at each timestep $t$, we generate $y_t$ (for $t=1, \ldots, T$) from $y_0$ using the same $\epsilon_t$ as (\ref{eq9}) following a previous work~\cite{song2022solving}:
\begin{equation}
    y_t = \sqrt{\bar{\alpha}_t} y_0 + \sqrt{1-\bar{\alpha}_t}\epsilon_t. \label{eq10} 
\end{equation}
By substituting (\ref{eq8}) and (\ref{eq9}) into (\ref{eq10}) and setting $\bar{\alpha}_0=1$, we obtain the following equation for $t=0, \ldots, T$:
\begin{equation}
    y_t = x_t + \sqrt{\bar{\alpha}_t}n. \label{eq11} 
\end{equation}
Eq. (\ref{eq11}) indicates that at each timestep $t$, there is a new denoising problem with noise $\sqrt{\bar{\alpha}_t}n$. Since $\bar{\alpha}_t$ decreases as $t$ increases and $\bar{\alpha}_T\approx0$~\cite{ho2020denoising}, we can use ${\widehat{x}_T=y_T}$ as an initialization. Then, for $t=T, \ldots, 1$, we can incorporate diffusion priors $p_\theta\left(x_{t-1} \mid x_t\right)$ into a MAP framework~\cite{zoran2011learning} to solve $x_{t-1}$:
\begin{equation}
    \begin{split}
        \widehat{x}_{t-1} &= \arg\min _{x_{t-1}} [\|x_{t-1}-y_{t-1}\|^2 \\
        &-\lambda_{t-1}\log p_\theta\left(x_{t-1} \mid \widehat{x}_t\right)], \label{eq13}
    \end{split}
\end{equation}
where $\lambda_{t}$ is a coefficient to trade off likelihood and prior at timestep $t$. In the reverse process of diffusion models, $p_\theta\left(x_{t-1} \mid \widehat{x}_t\right)$ follows a Gaussian distribution with mean $\mu_\theta\left(\widehat{x}_t, t\right)$ and diagonal covariance matrix $\sigma^2_{t}I$. When $t>1$, $\sigma_{t}>0$~\cite{ho2020denoising}, so we have:
\begin{equation}
    \begin{split}
        \widehat{x}_{t-1} &= \arg\min _{x_{t-1}}[\|x_{t-1}-y_{t-1}\|^2 \\
        &+\frac{\lambda_{t-1}}{\sigma_{t}} \|x_{t-1}-\mu_\theta\left(\widehat{x}_t, t\right)\|^2], \quad t>1. \label{eq14}
    \end{split}
\end{equation}
When $t=1$, $\sigma_{t}=0$, indicating that the final generation step of diffusion is deterministic:
\begin{equation}
    \widehat{x}_{t-1} = \mu_\theta\left(\widehat{x}_t, t\right), \quad t=1. \label{eq15}
\end{equation}
Eq. (\ref{eq14}) is a convex optimization problem that has a closed-form solution. By solving (\ref{eq14}) and combining it with (\ref{eq15}), we can derive the solution for $\widehat{x}_{t-1}$ in each iteration as follows:
\begin{equation}
    \label{eq16}
    \widehat{x}_{t-1}=\left\{
    \begin{aligned}
        &\frac{y_{t-1}+\lambda_{t-1} \mu_\theta\left(\widehat{x}_t, t\right)/\sigma_{t}}{1+\lambda_{t-1}/\sigma_{t}}, &t>1\\
        &\mu_\theta\left(\widehat{x}_t, t\right), & t=1
    \end{aligned}.
    \right.
\end{equation}
Similarly, if the pre-trained diffusion model is conditional on $c$, the solution becomes:
\begin{equation}
    \label{eq17}
    \widehat{x}_{t-1}^c=\left\{
    \begin{aligned}
        &\frac{y_{t-1}+\lambda_{t-1} \mu_\theta\left(\widehat{x}_t, t, c\right)/\sigma_{t}}{1+\lambda_{t-1}/\sigma_{t}}, &t>1\\
        &\mu_\theta\left(\widehat{x}_t, t, c\right), & t=1
    \end{aligned}.
    \right.
\end{equation}
Eq. (\ref{eq16}) and (\ref{eq17}) illustrate the denoising process using a pre-trained unconditional or conditional diffusion model. At each timestep, $y_{t-1}$ from (\ref{eq10}) helps correct the diffusion model's outputs, ensuring the accuracy of the generated images. The detailed algorithm is shown in Alg. \ref{alg1}.

Given a low-dose CT image $y_0^{hr}$ and two pre-trained diffusion models with parameters $\{\theta^{lr}, \{\bar{\alpha}_t^{lr}\}_{t=1,\ldots,T}\}$ and $\{\theta^{sr}, \{\bar{\alpha}_t^{sr}\}_{t=1,\ldots,T}\}$, we first obtain the low-resolution low-dose CT image $y_0^{lr}$ by $k\times$ downsampling. Next, we input $y_0^{lr}$, $c=None$, $\theta^{lr}$, $\{\bar{\alpha}_t^{lr}\}_{t=1,\ldots,T}$, $\{\lambda_t^{lr}\}_{t=1,\ldots,T-1}$ into Alg. \ref{alg1} to obtain the denoised low-resolution CT image $\widehat{x}_{0}^{lr}$, as shown in Fig. \ref{fig2}(a). Then, we input $y_0^{hr}$, $\widehat{x}_{0}^{lr}$, $\theta^{sr}$, $\{\bar{\alpha}_t^{sr}\}_{t=1,\ldots,T}$, $\{\lambda_t^{sr}\}_{t=1,\ldots,T-1}$ (where $\widehat{x}_{0}^{lr}$ is used as a condition) into Alg. \ref{alg1} to obtain the high-resolution denoised CT image $\widehat{x}_{0}^{hr}$, as shown in Fig. \ref{fig2}(b). The blue blocks in Fig. \ref{fig2} represent one iteration of our denoising algorithms. 

\begin{algorithm}[t]
	\caption{Image denoising with a pre-trained diffusion model}
	\label{alg1}
	\begin{algorithmic}[1]
		\Require $y_0$, $c$, $\theta$, $\{\bar{\alpha}_t\}_{t=1,\ldots,T}$, $\{\lambda_t\}_{t=1,\ldots,T-1}$.  
		\Ensure $\widehat{x}_{0}$.
		\State Initialize $t=T$. 
		\State Sample $\epsilon_T$ from $\mathcal{N}\left(0,I\right)$;
		\State $y_T=\sqrt{\bar{\alpha}_T} y_0+\sqrt{1-\bar{\alpha}_T}\epsilon_T$;
		\State $\widehat{x}_T=y_T$.
		\While{$t > 1$} 
		\State Sample $\epsilon_t$ from $\mathcal{N}\left(0,I\right)$;
		\State $y_{t-1}=\sqrt{\bar{\alpha}_{t-1}} y_0+\sqrt{1-\bar{\alpha}_{t-1}}\epsilon_{t-1}$;
		\If{$c == None$}
		\State Update $\widehat{x}_{t-1}$ according to (\ref{eq16});
		\Else
		\State Update $\widehat{x}_{t-1}$ according to (\ref{eq17});
		\EndIf
		\State $t=t-1$;
		\EndWhile
		\State $\widehat{x}_0=\mu_\theta\left(\widehat{x}_1, 1\right)$.
	\end{algorithmic}
\end{algorithm}
\subsubsection{Adaptive Refinement of $\lambda$} 

In Alg. \ref{alg1}, the hyper-parameters $\{\lambda_t\}_{t=1,\ldots,T-1}$ are used to balance the trade-off between likelihood and prior. Empirically, a lower noise level requires a smaller values for $\lambda$. According to (\ref{eq11}), the noise level at timestep $t$ is given by $\sqrt{\bar{\alpha}_t}n$, where $\bar{\alpha}_t$ decreases from $1$ to nearly $0$ as $t$ increases from $0$ to $T$. Therefore, we tentatively assign $\lambda_t$ as the product of $\bar{\alpha}_t$ and a constant $\lambda_0$,

\begin{equation}
	\lambda_t=\lambda_0\cdot\sqrt{\bar{\alpha}_t}. \label{eq22}
\end{equation}

We refer to this strategy as \textbf{ConsLam}. However, different low-dose CT slices often exhibit varying noise levels, making it challenging for a fixed $\lambda_0$ to achieve the optimal denoising result for each image. To adaptively denoise low-dose CT images with different noise levels, we first propose to estimate the noise level of a single low-dose CT image. 

Directly estimating the noise level from a single low-dose CT image is difficult due to the non-analytical nature of low-dose CT noise. Therefore, we first denoise the low-dose CT image using a fixed $\lambda_0$ to obtain an initial denoising result $\widehat{x}_0$. The estimated noise is then calculated as the absolute difference between the original image $y_0$ and the denoised image $\widehat{x}_0$,

\begin{equation}
	\widehat{n} = |y_0-\widehat{x}_0|.
\end{equation}

Although the estimated $\widehat{n}$ obtained using a fixed $\lambda_0$ may not be optimal, it still provides a reliable representation of the noise in the input low-dose CT image $y_0$. We adjust $\lambda_0$ based on $\widehat{n}$ and then resume the denoising algorithm from a certain intermediate step to achieve more accurate denoising. Specifically, we propose two strategies for refining $\lambda_0$.

The first strategy involves using the standard deviation of the estimated noise as the noise level to dynamically determine $\lambda_0$. This is expressed as $\lambda_0^{ada}$, 
\begin{equation}
	\lambda_0^{ada} = a \cdot std(\widehat{n}) + b, \label{eq24}
\end{equation}
where $\text{std}(\cdot)$ denotes the standard deviation of all pixels in an image, and $a$ and $b$ are manually selected parameters. We refer to this strategy as \textbf{AdaLam-\uppercase\expandafter{\romannumeral1}}.

The second strategy takes into consideration that different regions of a low-dose CT image exhibit various levels of noise. To address this, we introduce an uneven $\lambda_0$ across different positions, resulting in a matrix $\Lambda_0^{ada}$,
\begin{equation}
    \Lambda_0^{ada} = c \cdot \widehat{n}, \label{eq25_}
\end{equation}
where $c$ is a manually selected coefficient. To incorporate the matrix $\Lambda_0^{ada}$ into Alg. \ref{alg1} in place of the scalar $\lambda_0$, we perform the Hadamard product. We refer to the second strategy as \textbf{AdaLam-\uppercase\expandafter{\romannumeral2}}. In Section \uppercase\expandafter{\romannumeral4}.B, we validate the effectiveness of using each of these two strategies separately, as well as using them simultaneously. 

\begin{figure}[t]
	\centering\includegraphics[width=0.45\textwidth]{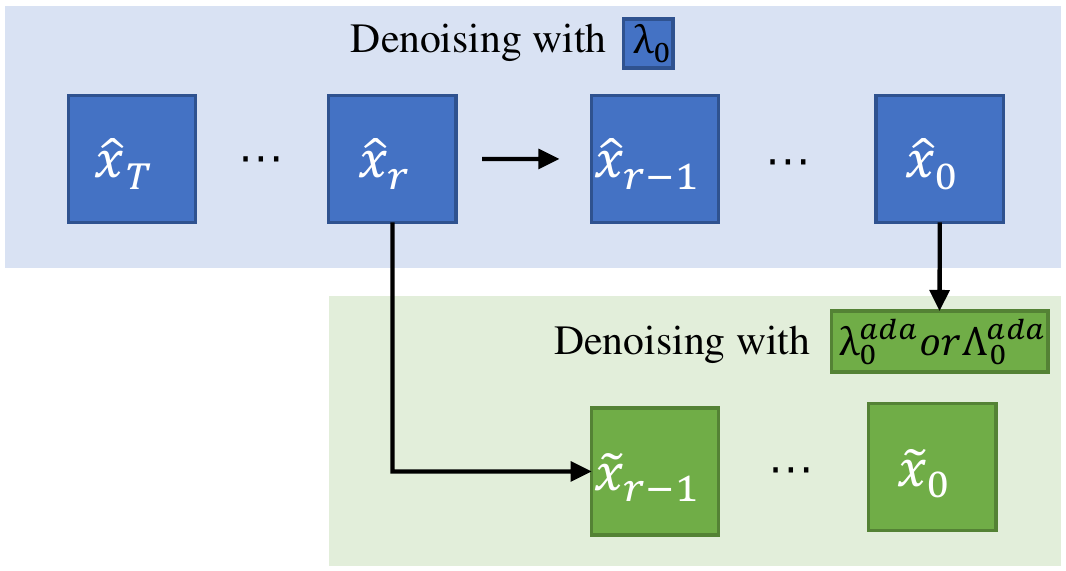}
	\caption{Refinement of $\lambda$ after coarse denoising. Based on the denoising result with a fixed $\lambda_0$, we can derive $\lambda_0^{ada}$ or $\Lambda_0^{ada}$ for each low-dose CT image. Subsequently, at timestep $r$, we can resume and perform more accurate denoising using $\lambda_0^{ada}$ or $\Lambda_0^{ada}$.} \label{fig3}
\end{figure}

Fig. \ref{fig3} illustrates the process of refining $\lambda_0$ in detail. It is worth noting that resuming at an larger step can yield a more accurate $\lambda_0^{ada}$ or $\Lambda_0^{ada}$ from $\widehat{n}$. In theory, resuming at timestep $T$ would be the optimal choice. However, this approach would double the computational time of the algorithm. Therefore, to balance accuracy and computation time, we choose to roll back only a few steps. In our experiments, we set the roll-back step as 3, resuming from $\widehat{x}_{3}$, which only increases the inference time by approximately 10\%. It is worth noting that we only employ the adaptive refinement strategies of $\lambda$ in the second phase of our denoising algorithm since a constant $\lambda_0$ is sufficient for low-resolution image denoising.

\subsubsection{Acceleration} 

The time-consuming nature of diffusion models, which rely on an iterative generation process, poses a significant limitation. The same issue affects our denoising algorithm based on diffusion models. To address this, we adopt the approach proposed by DDIM~\cite{song2020denoising} to expedite the generation process of diffusion models through interval sampling. Specifically, we extract a sub-sequence $\tau=[\tau_1, \ldots, \tau_S]$ of length $S$ from $[1, \ldots, T]$, and then perform sampling from $x_{\tau_s}$ to $x_{\tau_{s-1}}$. For our denoising algorithm, the number of iterations decreases according to the following relation:

\begin{equation}
	\widehat{x}_{\tau_{s-1}}=\frac{y_{\tau_{s-1}}+\lambda_{\tau_{s-1}}\mu_\theta\left(\widehat{x}_{\tau_s}, \tau_s\right)/\sigma_{\tau_{s}}}{1+\lambda_{\tau_{s-1}}/\sigma_{\tau_{s}}},\quad s>1. \label{eq25}
\end{equation}         

For the accelerated algorithms, Eq. (\ref{eq25}) replaces (\ref{eq16}) when $t>1$. In the case of a conditional diffusion model, the implementation is similar, substituting $\mu_\theta\left(\widehat{x}_{\tau_s}, \tau_s\right)$ with $\mu_\theta\left(\widehat{x}_{\tau_s}, \tau_s, c\right)$.  

In typical generative tasks, the timesteps of accelerated sampling are often uniformly reduced. However, in this paper, we propose to adopt an non-uniform strategy to extract the sub-sequence from the total of 2000 timesteps. Specifically, within the first 500 timesteps, we extract one every 20 timesteps. In the subsequent 1500 timesteps, we extract one every 500 timesteps. Finally, we include the final timestep $T$ to ensure the validity of the initial conditions of our algorithm, i.e., $\widehat{x}_T=y_T$. Consequently, we set $\tau=[1, 21, 41,\ldots,501, 1001, 1501, 2000]$ and reduce the total number of iterations from 2000 to 29. This non-uniform sampling approach is designed because the likelihood term plays a more prominent role in the smaller timesteps, and denser sampling enables better utilization of the low-dose CT input, resulting in more accurate denoising results.
 \begin{figure}[t]
	\centering\includegraphics[width=0.5\textwidth]{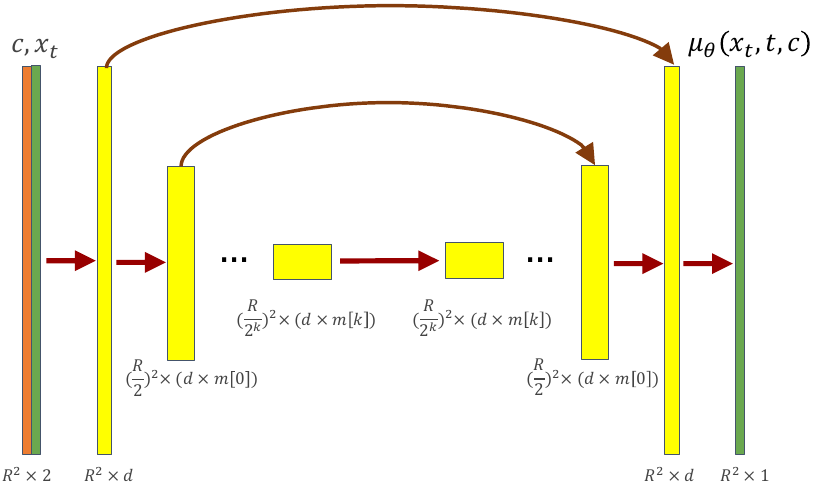}
	\caption{U-Net architecture used in diffusion models, where $R$ denotes the image size, $d$ denotes the initial channel dimension, and $\{m[1],\ldots,m[k]\}$ denotes the channel multipliers.} \label{fig4}
\end{figure}

\section{Implementation Details}
\subsection{Dataset}
We pre-train two diffusion models respectively on abdomen and chest normal-dose CT images of 10 patients from the Low-dose CT and Projection Data (LDCT-PD) dataset~\cite{clark2013cancer,mccollough2020low}. For abdomen and chest CT images, the high-resolution image size is both $512\times512$, while the low-resolution image sizes are $128\times128$ and $256\times256$ respectively. To evaluate the denoising performance of our proposed Dn-Dp algorithm, we employed quarter-dose abdomen CT images of 4 patients from the 2016 AAPM Grand Challenge dataset~\cite{chen2017low} and 10\%-dose chest CT images of 4 patients from the LDCT-PD dataset, ensuring that these images were distinct from the training data. 

\subsection{Network Architecture and Parameters}
For our diffusion models, we adopt the U-Net architecture from SR3~\cite{saharia2022image}, while the residual blocks are sourced from BigGAN~\cite{brock2018large}. In the case of conditional diffusion models, the low-resolution image is upsampled using bicubic interpolation and subsequently concatenated with the previous output of the U-Net along the channel dimension. Table \ref{tab1} provides the specifics for the number of initial channel dimensions (Dim $d$), the channel multipliers (Muls $\{m[1],\ldots,m[k]\}$), the number of residual blocks (ResBlocks), the resolution at which a self-attention residual block is employed (Res-Attn), and the dropout rate (Dropout). The architecture of the conditional U-Net used in our diffusion models is depicted in Fig. \ref{fig4}. To obtain the unconditional version, the condition image $c$ is simply removed from the model.

\begin{table}[thb]\centering
	\caption{Detailed Parameters of the U-Nets Utilized in This Study.}
	\label{tab1}
	\resizebox{0.48\textwidth}{!}{
		\large
		\begin{tabular}{*{7}{c}}
			\toprule
			\multirow{2}{*}{Dataset}& \multirow{2}{*}{Resolution} &  {Dim} & {Muls} & \multirow{2}{*}{ResBlocks} & \multirow{2}{*}{Res-Attn} & \multirow{2}{*}{Dropout} \\
			~ & ~ &  $d$ & $\{m[1],\ldots,m[k]\}$ &  ~ & ~ &  ~  \\
			\midrule
			\multirow{2}{*}{Abodmen} & $128\times 128$ & 64 & \{1,2,4,8,8\}& 2 & 16 & 0.2 \\
			~ & $128\times 128\rightarrow 512\times 512$ & 64 & \{1,2,4,8,16,32,32\} & 2 & / & 0 \\
			\midrule
			\multirow{2}{*}{Chest} & $256\times 256$ & 64 & \{1,2,4,8,8\} & 4 & 16 & 0.2 \\
			~ & $256\times 256\rightarrow 512\times 512$ & 64 & \{1,2,4,8,8,16,16\} & 5 & / & 0 \\
			\bottomrule
		\end{tabular}
	}
\end{table}

For the pre-trained two cascaded diffusion models, we set the number of timesteps, denoted as $T$, to be 2000. In the forward processes, we set $\beta_1=1\times10^{-6}$, $\beta_T=1\times10^{-2}$, and $\beta_i(1<i<T)$ to increase linearly at the intermediate timesteps. The proposed denoising algorithm involves several hyper-parameters, namely $\lambda_0$, $a$, $b$, and $c$ in (\ref{eq22}), (\ref{eq24}), and (\ref{eq25_}), which are manually adjusted respectively for denoising abdomen and chest CT images. We have provided the selected parameters in Table \ref{tab2}. 

\begin{table}[thb]\centering
	\caption{Hyper-parameters in Denoising Algorithms}
	\label{tab2}
	\resizebox{0.48\textwidth}{!}{
		\large
		\begin{tabular}{*{6}{c}}
			\toprule
			Dataset& Resolution &$\lambda_0$ &  $a$ & $b$ & $c$  \\
			\midrule
			\multirow{2}{*}{Abodmen} & $128\times 128$ & 0.002 & /& / & / \\
			~ & $128\times 128\rightarrow 512\times 512$ & 0.0075 & 1.5 & -0.01 & 0.3 \\
			\midrule
			\multirow{2}{*}{Chest} & $256\times 256$ & 0.03 & / & / & / \\
			~ & $256\times 256\rightarrow 512\times 512$ & 0.09 & 3.5 & -0.09 & 1.2 \\
			\bottomrule
		\end{tabular}
	}
\end{table}

 Moreover, the varying noise introduced during the forward process yields denoised results that exhibit consistent distributions but differ in pixel values. To enhance performance, we calculate the average of 10 denoising results~\cite{kawar2021snips}.

\section{Experiments}
\subsection{Normal-dose CT Image Generation}
\begin{figure}[htbp]
	\centering\includegraphics[width=0.45\textwidth]{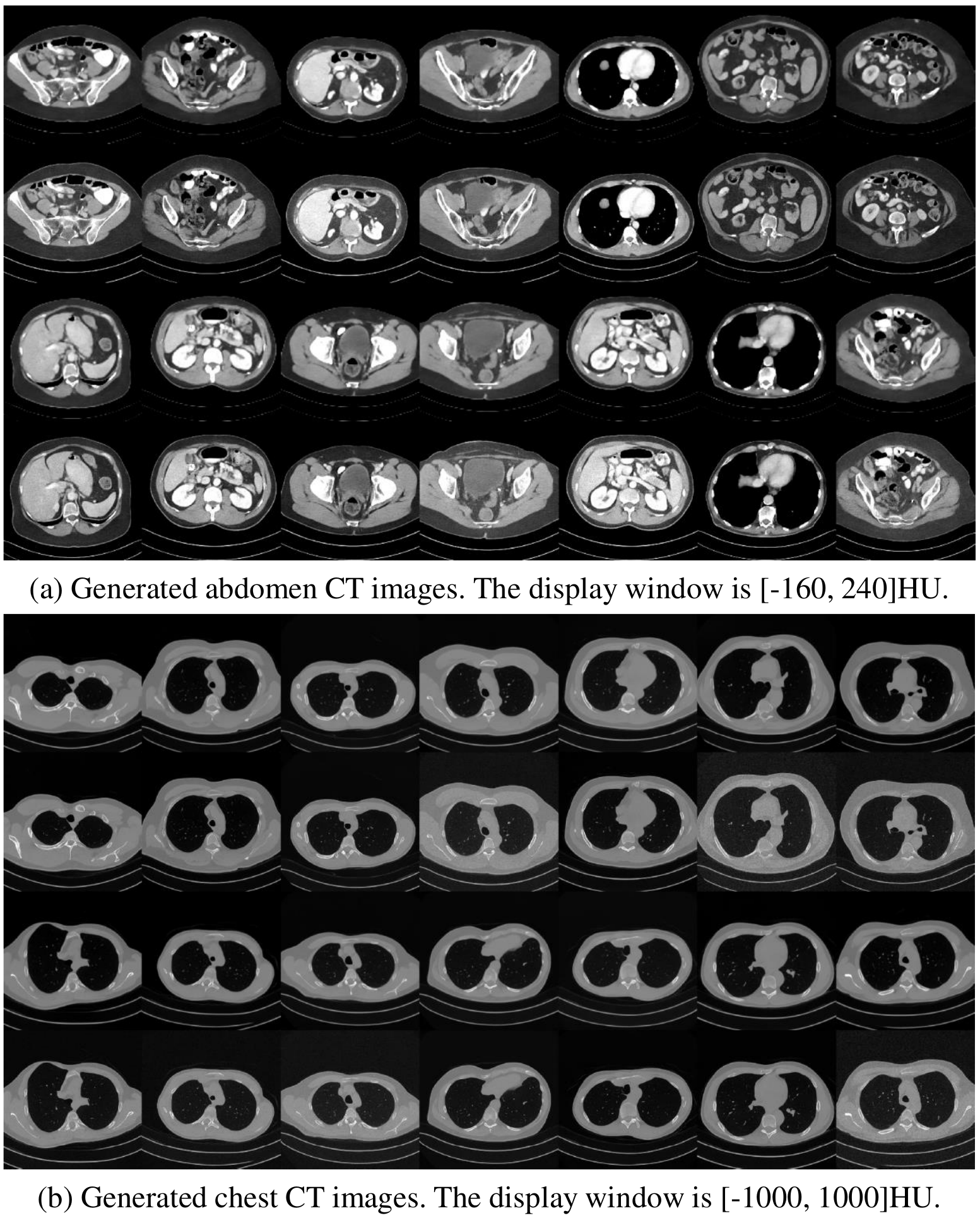}
	\caption{Normal-Dose CT images generated by pretrained cascaded diffusion models. The odd rows display low-resolution images, while the even rows show the corresponding high-resolution images.} \label{fig5}
\end{figure}

Fig. \ref{fig5} showcases the high-quality normal-dose CT images generated by our pre-trained cascaded diffusion models. These generated images exhibit remarkable realism and diversity, indicating that our well-trained diffusion model possesses a strong prior understanding of normal-dose CT images. 

\begin{figure*}[t]
	\centering\includegraphics[width=0.95\textwidth]{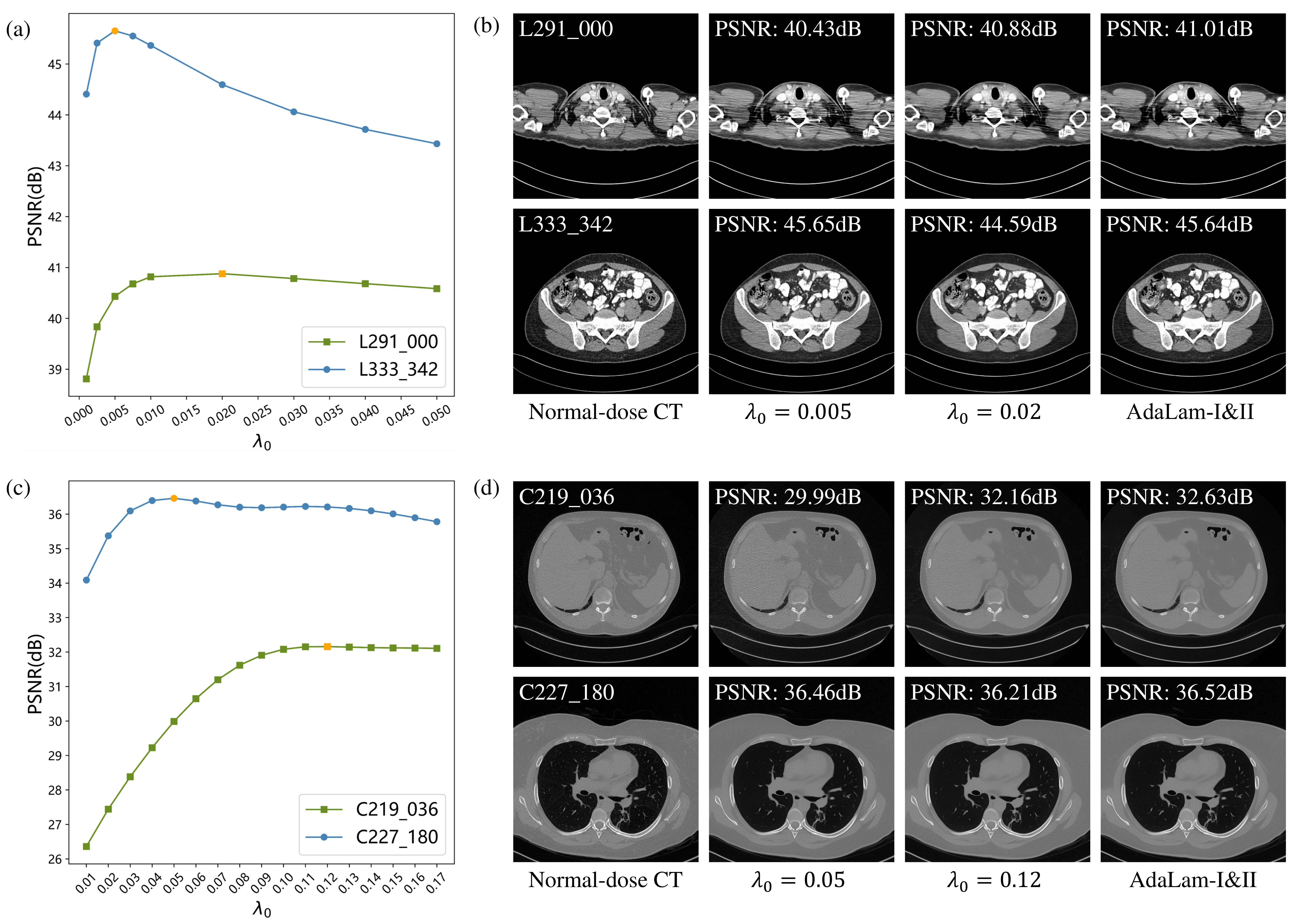}
	\caption{(a) and (c) depict the relationship between denoised image PSNR and the constant $\lambda_0$ for two abdomen low-dose CT images and two chest low-dose CT images. (b) and (d) showcase visual results of the four selected low-dose CT images with optimal constant $\lambda_0$ and adaptive $\lambda_0^{ada}\&\Lambda_0^{ada}$. The abdomen and chest CT images are respectively displayed with the window [-160, 240] HU and [-1000, 1000] HU. PSNRs are calculated using the window [-1024, 3072] HU.} \label{fig8}
\end{figure*}
\subsection{Ablation Study of $\lambda$}
\begin{table}[thb]\centering
	\caption{Quantitative Results of Different $\lambda$ Strategies}
	\label{tab3}
	\resizebox{0.48\textwidth}{!}{
		\large
		\begin{tabular}{*{6}{c}}
			\toprule
			Dataset & Metrics& ConsLam &AdaLam-\uppercase\expandafter{\romannumeral1} &  AdaLam-\uppercase\expandafter{\romannumeral2} & AdaLam-\uppercase\expandafter{\romannumeral1}\&\uppercase\expandafter{\romannumeral2}  \\
			\midrule
			\multirow{2}{*}{Abodmen}&PSNR & 44.80 & \textbf{44.90} & 44.86 &\textbf{44.90} \\
			~ &SSIM& 0.971 & 0.972 & 0.972 & \textbf{0.973} \\
			\midrule
			\multirow{2}{*}{Chest} &PSNR& 34.79 & 34.92 & 35.20 & \textbf{35.29} \\
			~ & SSIM & 0.791 & 0.796 & 0.806 & \textbf{0.811} \\
			\bottomrule
		\end{tabular}
	}
\end{table} 

In this subsection, we evaluate the effectiveness of our proposed refinement strategy for $\lambda$ as presented in (\ref{eq24}) and (\ref{eq25_}) including AdaLam-\uppercase\expandafter{\romannumeral1}, AdaLam-\uppercase\expandafter{\romannumeral2}, and AdaLam-\uppercase\expandafter{\romannumeral1}\&\uppercase\expandafter{\romannumeral2}, which denotes using them simultaneously. To conduct the ablation studies, we uniformly extract 80 CT slices from the abdomen and chest data of four patients, resulting in two mini-datasets. Table \ref{tab3} presents the statistical metrics of the denoising results. The metrics are calculated within the window [-1024, 3072] HU. From the results in Table \ref{tab3}, it can be observed that for abdomen CT image denoising, both AdaLam-\uppercase\expandafter{\romannumeral1} and AdaLam-\uppercase\expandafter{\romannumeral2} yield an improvement of approximately 0.1dB in mean PSNR (peak signal-to-noise ratio), and their combined usage produces equally good results. On the other hand, for chest CT image denoising, where the noise level is higher, AdaLam-\uppercase\expandafter{\romannumeral1} and AdaLam-\uppercase\expandafter{\romannumeral2} exhibit more significant enhancements in terms of PSNR and SSIM (structural similarity index measure) compared to the ConsLam. Furthermore, utilizing AdaLam-\uppercase\expandafter{\romannumeral1}\&\uppercase\expandafter{\romannumeral2} for chest CT images leads to superior results compared to using either one alone.
\begin{table*}[t]\centering
	\caption{Quantitative Results of Various Methods on CT Images of 4 Patients for Abdomen and Chest CT Image Denoising, respectively. Bold indicates the best index among unsupervised methods, while asterisk indicates the best index among all methods.}
	\label{tab4}
	\resizebox{\textwidth}{!}{
		\large
		\begin{tabular}{*{11}{c}}
			\toprule
			\multicolumn{3}{c}{} & \multicolumn{4}{c}{Unsupervised Methods} & \multicolumn{2}{c}{Supervised Methods} & \multicolumn{2}{c}{Ours} \\
			\midrule
			Dataset & Metrics& Low-dose & BM3D~\cite{dabov2007image} &  NLM~\cite{buades2005non} & LIR-IR~\cite{du2020learning} & Noise2Sim~\cite{niu2022noise} & RED-CNN~\cite{chen2017low}  & CTformer~\cite{wang2023ctformer} & Dn-Dp (Single) &  Dn-Dp (Average) \\
			\midrule
			\multirow{2}{*}{Abodmen}&PSNR & 40.41 & 43.14 & 42.85 & 40.63 & 44.63 & 44.96 & $45.07^*$  & 44.81 & \textbf{45.02} \\
			~ &SSIM& 0.922 & 0.955 & 0.954 & 0.951 & 0.972 & 0.973 & $0.974^*$  & 0.971 &\textbf{0.973} \\
			\midrule
			\multirow{2}{*}{Chest} &PSNR& 27.55 & 34.37 & 33.73 & 33.73 & 34.50 & $37.07^*$ & 35.38  & 35.25 &\textbf{35.35} \\
			~ & SSIM & 0.532 & 0.812 & 0.795 & 0.801 & 0.808 & $0.879^*$ & 0.834  & 0.809 & \textbf{0.813} \\
			\bottomrule
		\end{tabular}
	}
\end{table*} 

The utilization of adaptive refinement strategies leads to enhanced denoising outcomes due to the varying noise levels observed in different low-dose CT images. Generally, low-dose CT images characterized by lower noise levels are better suited for a smaller $\lambda_0$, whereas those with higher noise levels benefit from a larger $\lambda_0$. In Fig. \ref{fig8}(a) and \ref{fig8}(c), the relationship between the denoised image PSNR and the constant $\lambda_0$ is plotted for two abdomen low-dose CT images and two chest low-dose CT images, respectively. It is evident that employing the ConsLam results in different CT images achieving optimal denoising outcomes with distinct $\lambda_0$ values. By calculating the adaptive $\lambda_0^{ada}$ and $\Lambda_0^{ada}$, and subsequently resuming the denoising algorithm for a few iterations, we can achieve better denoising results for low-dose CT images with different noise levels. As depicted in Fig. \ref{fig8}(b) and \ref{fig8}(d), simultaneous adaptation of AdaLam-\uppercase\expandafter{\romannumeral1} and AdaLam-\uppercase\expandafter{\romannumeral2} yields denoising performance that either surpasses or approximates the best achievable results obtained using the ConsLam. This elucidates why the adoption of adaptive $\lambda$ strategies enhances the algorithm's performance across the entire dataset, as demonstrated in Table \ref{tab3}.


\subsection{Comparison of Denoising Performance}

Based on the ablation studies of $\lambda$, we have determined that the simultaneous usage of both AdaLam-\uppercase\expandafter{\romannumeral1} and AdaLam-\uppercase\expandafter{\romannumeral2} yields the best performance for our denoising algorithms. In the following experiments, we employ AdaLam-\uppercase\expandafter{\romannumeral1}\&\uppercase\expandafter{\romannumeral2} for Dn-Dp. Additionally, we evaluate two versions of Dn-Dp: Dn-Dp (Single), which applies denoising once to the input image, and Dn-Dp (Average), which generates multiple denoised results for the same input image and averages them. The latter approach can improve performance but requires more computational time.

We compare our method with various low-dose CT image denoising algorithms, including two classic non-learning methods: Block matching 3D (BM3D)~\cite{dabov2007image} and Non-local means (NLM)~\cite{buades2005non}. Moreover, we consider two supervised learning-based neural networks that have achieved excellent results in low-dose CT image denoising: RED-CNN~\cite{chen2017low} and CTformer~\cite{wang2023ctformer}, with RED-CNN based on CNNs and CTformer being a transformer-based model. Since our method operates in a fully unsupervised manner, we also compare it with two unsupervised learning-based methods: LIR-IR~\cite{du2020learning} and Noise2Sim~\cite{niu2022noise}, with the latter being the state-of-the-art unsupervised low-dose CT image denoising method.

Table \ref{tab4} presents the quantitative denoising results of the different methods, with the metrics calculated using the window [-1024, 3072] HU. Our proposed methods demonstrate competitive quantitative results in terms of PSNR and SSIM, as shown in Table \ref{tab4}. Specifically, on the abdomen CT dataset, our methods outperform all of the unsupervised methods. Notably, our method achieves a higher PSNR than the classic supervised deep learning-based methods, RED-CNN, and approaches the performance of the latest transformer-based network, CTformer. This result is surprising since supervised networks trained with MSE (Mean Square Error) loss typically prioritize higher PSNR values. However, for the chest CT dataset, where low-dose CT images exhibit more pronounced noise, CNN-based supervised neural networks such as RED-CNN demonstrate the best performance. Due to the higher noise levels, the performance gap between unsupervised and supervised algorithms becomes more notable. Nevertheless, our methods remain the best-performing among unsupervised methods, surpassing LIR-IR and Noise2Sim.
\begin{figure*}[thb]
	\centering\includegraphics[width=0.98\textwidth]{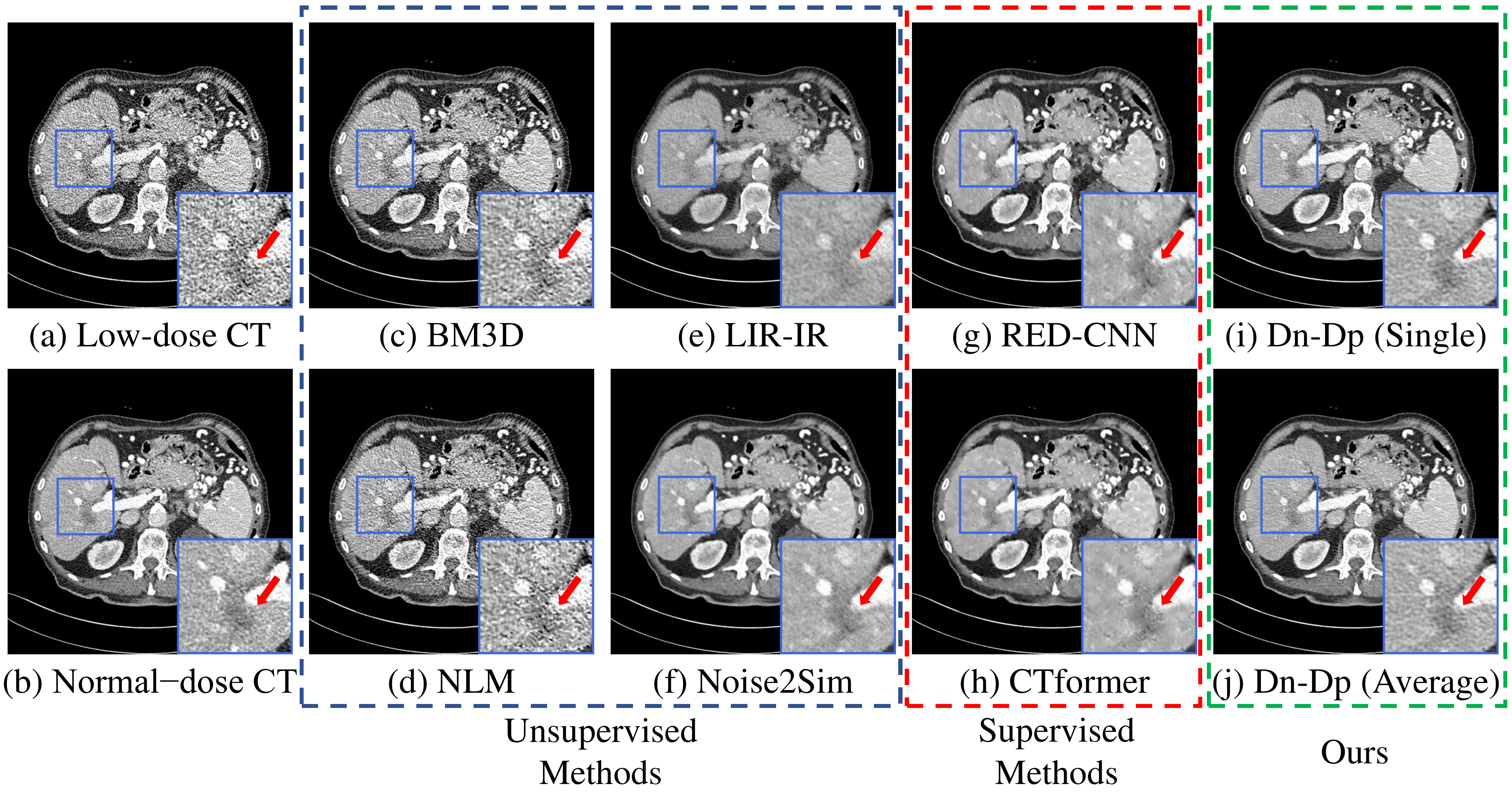}
	\caption{Visual comparison of abdomen CT images.} \label{fig9}
\end{figure*}
\begin{figure*}[thb]
	\centering\includegraphics[width=0.98\textwidth]{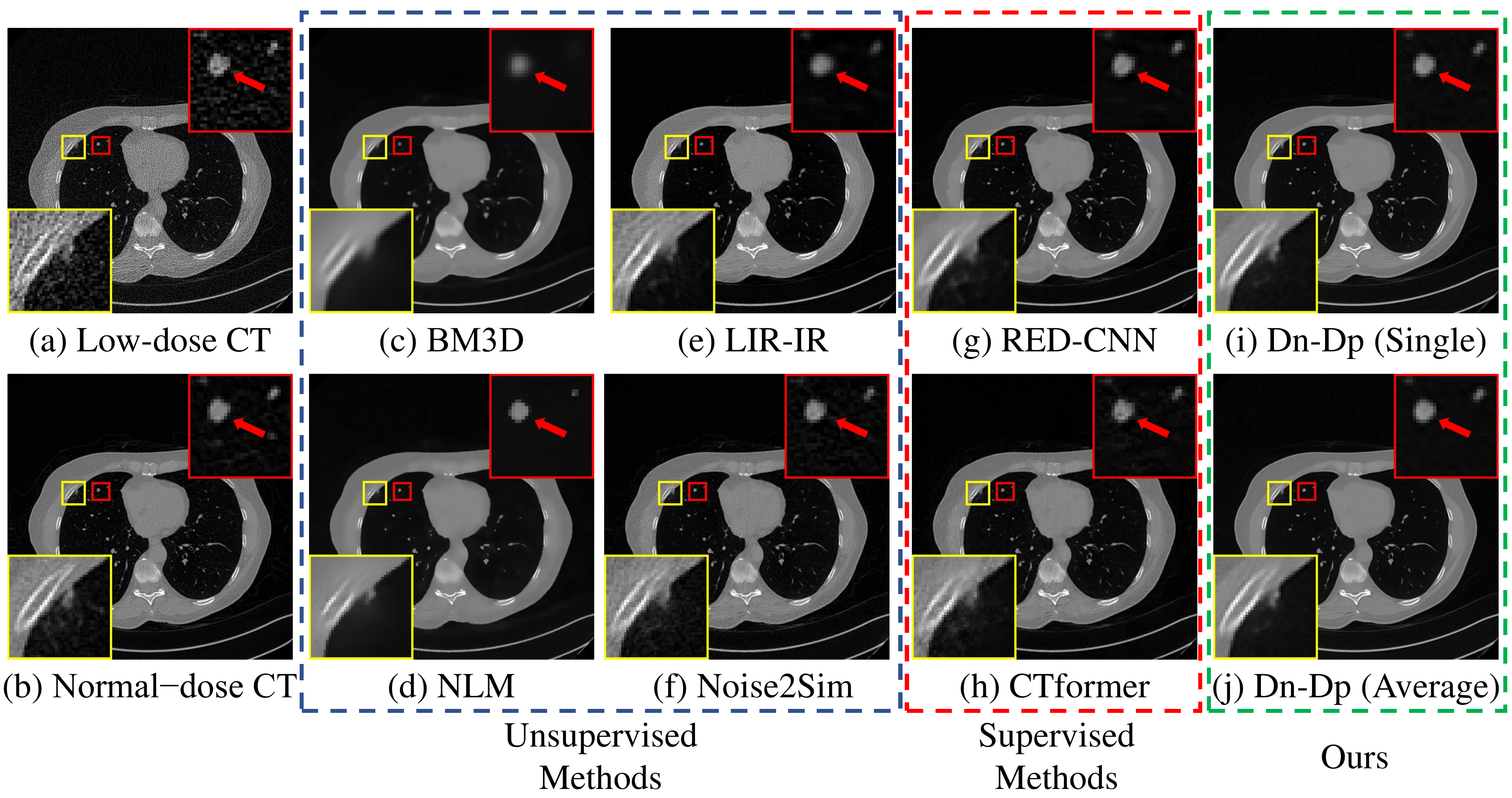}
	\caption{Visual comparison of chest CT images.} \label{fig10}
\end{figure*}
Fig. \ref{fig9} illustrates the denoising outcomes for an abdomen CT image. It is evident that our method produces results that closely resemble the visual quality of the normal-dose CT image. By comparing (c) and (d), it can be observed that traditional methods fail to effectively eliminate the noise present in low-dose CT. On the other hand, (e) and (f) demonstrate that unsupervised deep learning-based methods yield better results than traditional methods but still tend to lose some texture details found in the normal-dose CT image due to their end-to-end nature. This issue becomes more pronounced in (g) and (h) for supervised deep learning algorithms, resulting in noticeable oversmoothing. In contrast, our method, which utilizes the MAP framework combining prior and likelihood, achieves visually more realistic results.

In Fig. \ref{fig10}, the visual results of a chest CT image are presented. Within the yellow region of interest (ROI), the ability of different algorithms to recover bone structures is compared. Due to the higher noise levels in the 10\% dose chest CT, it can be observed that even supervised algorithms lose some contour information. However, our methods strive to preserve the contour as much as possible during denoising. Within the red ROI, the arrows indicate a solid non-calcified pulmonary nodule. Denoising with BM3D, NLM, and LIR-IR leads to blurring of the nodule, which can hinder clinical diagnosis. In contrast, our methods and the supervised methods effectively maintain the shape of the nodule.

\subsection{Discussion}

While our method has achieved excellent results in terms of visual effects and quantitative metrics, there are still some limitations worth noting. The main limitations of our methods lie in the size of the model parameters and the computational time required. To train the diffusion model for generating high-resolution normal-dose CT images, a cascaded generation approach and a significantly large model size are necessary. Despite the use of an acceleration algorithm that reduces the number of iterations from 2000 to 29, the inference time for denoising a single image remains much higher compared to end-to-end networks. Table \ref{tab6} presents the parameter sizes of the models used and the corresponding inference times required for denoising using these models. Here, Time(1) and Times(10) respectively represent the inference time for denoising a single image and 10 images in parallel using a single NVIDIA GeForce RTX 3090 graphics card.
\begin{table}[htb]\centering
	\caption{Parameter Size of Diffusion Models and Inference Time of Our Denoising Methods}
	\label{tab6}
	\resizebox{0.48\textwidth}{!}{
		\large
		\begin{tabular}{*{5}{c}}
			\toprule
			Dataset & Resolution &  Parameters & Time(1) & Times (10)  \\
			\midrule
			\multirow{2}{*}{Abodmen} & $128\times 128$ & 373M &  1.56s & 2.01s  \\
			~ & $128\times 128\rightarrow 512\times 512$ & 5.49G & 5.83s & 34.67s \\
			\midrule
			\multirow{2}{*}{Chest} & $256\times 256$ & 605M & 2.56s & 10.36s  \\
			~ & $256\times 256\rightarrow 512\times 512$ & 2.94G & 8.05s & 56.20s \\
			\bottomrule
		\end{tabular}
	}
\end{table}

An alternative after denoising the low-resolution images with Dn-Dp is directly training a super-resolution neural network to acquire high-resolution clean images. Although the super-resolution network requires a supervised training, training pairs can be produced from only normal-dose CT images by downsampling. Using such a super-resolution network to replace the second stage of Dn-Dp will inevitably result in performance loss since it does not utilize the information from high-resolution low-dose CT images, it can save some inference time and provide reasonable denoising results. 

Furthermore, the reason we choose to use a cascaded manner for high-resolution image generation is to facilitate the pre-training of diffusion models. However, this approach may lead to a performance decrease because downsampling actually introduces further degradation to the images, resulting in some information loss. If we can train an unconditional diffusion model directly for high-resolution image generation, our algorithm is also effective for one-stage denoising, and the results may even be better. However, training such a high-resolution unconditional diffusion model often requires deeper networks, more training data, and longer training time.

\section{Conclusion}
In this paper, we propose Dn-Dp, a diffusion prior-based algorithm for denoising low-dose CT images. Dn-Dp is a completely unsupervised algorithm since it only requires training the network with normal-dose CT images. The algorithm involves training a cascaded diffusion model to generate normal-dose CT images. Subsequently, during the reverse process of the diffusion models, we iteratively solve multiple MAP estimation problems. This ensures a high likelihood between the generated images and the input low-dose CT images while maintaining good image quality. To address the varying noise levels in different low-dose CT images, we employ two adaptive strategies to adjust the coefficients $\lambda$, which balance the likelihood and prior terms in the MAP estimation. We also introduce a method to resume denoising at an intermediate stage with more accurate $\lambda$. Our methods demonstrate excellent results in terms of quantitative metrics and visual effects, particularly in denoising abdomen CT images at a quarter dose, surpassing even supervised methods in terms of PSNR and SSIM. Furthermore, we aim to extend the application of diffusion prior-based methods to CT reconstruction algorithms, starting from the sinogram domain.


\section*{Acknowledgment}
We express our gratitude to the creators of SR3~\cite{saharia2022image} from the github repository "Image-Super-Resolution-via-Iterative-Refinement" for providing the codes that facilitated the training of the cascaded diffusion models. We have made our codes and pre-trained models publicly available.

\bibliographystyle{IEEEtran}
\bibliography{IEEEabrv,tmi2023}

\end{document}